\documentclass[aps,pra,reprint, amsmath, amssymb,superscriptaddress,nofootinbib]{revtex4-1}

\usepackage{bm}
\usepackage[retainorgcmds]{IEEEtrantools}
\usepackage{graphicx}
\usepackage{mathrsfs}
\usepackage{amsmath}
\usepackage{amssymb}
\usepackage{color}
\usepackage{amsfonts}
\usepackage{times,txfonts}
\usepackage{nicefrac}
\usepackage[colorlinks=true,linkcolor=blue,urlcolor=blue,citecolor=blue,pdfusetitle]{hyperref}

\begin{document}

\title{Information bound for entropy production from the detailed fluctuation theorem}
\date{\today}
\author{Domingos S. P. Salazar}
\affiliation{Unidade de Educa\c c\~ao a Dist\^ancia e Tecnologia,
Universidade Federal Rural de Pernambuco,
52171-900 Recife, Pernambuco, Brazil}

\begin{abstract}
Fluctuation theorems impose fundamental bounds in the statistics of the entropy production, with the second law of thermodynamics being the most famous. Using information theory, we quantify the information of entropy production and find an upper tight bound as a function of its mean from the strong detailed fluctuation theorem. The bound is given in terms of a maximal distribution, a member of the exponential family with nonlinear argument. We show that the entropy produced by heat transfer using a bosonic mode at weak coupling reproduces the maximal distribution in a limiting case. The upper bound is extended to the continuous domain and verified for the heat transfer using a levitated nanoparticle. Finally, we show that a composition of qubit swap engines satisfies a particular case of the maximal distribution regardless of its size.
\end{abstract}
\maketitle{}

%\section{Introduction}

% {\bf \emph{Introduction -}}
{\bf \emph{Introduction -}} Fluctuation theorems (FTs) have far reaching consequences in nonequilibrium thermodynamics. As experiments probe smaller setups, entropy production $\Sigma$ is seen as a random variable. In this situation, FTs impose constraints in the distribution $P(\Sigma)$ by requiring that not only a positive entropy production is observed on average $\langle \Sigma \rangle\geq0$, but also quantifying its chance with respect to the time-reversed event \cite{Seifert2012,Campisi2011,Bustamante2005,Jarzynskia2008}.  

Among some variations of FTs \cite{Jarzynski1997,Jarzynski2000,Crooks1998,Gallavotti1995,Evans1993,Hanggi2015,Saito2008}, we focus our attention in the strong detailed fluctuation theorem (DFT),
\begin{equation}
\label{DFT}
    \frac{P(\Sigma)}{P(-\Sigma)}=e^{\Sigma},
\end{equation}
which results from time symmetric protocols in the framework of the exchange fluctuation theorem (EFT) \cite{Hasegawa2019,Timpanaro2019B,Evans2002,Merhav2010,Garcia2010,Cleuren2006,Seifert2005,Jarzynski2004a,Andrieux2009,Campisi2015}. In the EFT case, a set of charges $(\mathcal{Q}_1, ...,\mathcal{Q}_N)$ are observed in a finite time experiment and, with their respective affinities $A_i$, they satisfy the EFT, $P(\mathcal{Q}_1,\mathcal{Q}_2,...,\mathcal{Q}_N)/P(-\mathcal{Q}_1,-\mathcal{Q}_2,...,-\mathcal{Q}_N)=\exp(\sum_i A_i \mathcal{Q}_i)$.
The entropy production random variable is given by $\Sigma = \sum_i A_i \mathcal{Q}_i$. Focusing on the actual distribution of $\Sigma$ \cite{Hasegawa2019,Timpanaro2019B}, one defines $P(\Sigma):=\int \delta (\Sigma - \sum_i A_i\mathcal{Q}_i) P(\mathcal{Q}_1,...,\mathcal{Q}_N)d\mathcal{Q}_1...d\mathcal{Q}_N$, which satisfies (\ref{DFT}).

Although the DFT (\ref{DFT}) still leaves plenty of room for a variety of possible distributions $P(\Sigma)$, some fundamental bounds are imposed in their statistics \cite{Merhav2010,Timpanaro2019B,Hasegawa2019} as well as generic properties \cite{Neri2017,Pigolotti2017}. For instance, (\ref{DFT}) implies the integral fluctuation theorem $\langle e^{-\Sigma}\rangle=1$, which, in turn, results in the second law $\langle \Sigma \rangle \geq 0 $ from Jensen's inequality. In this case, if the second law is a fundamental bound derived from the FT, perhaps other bounds might also play important roles. 

Following this idea, other bounds were obtained recently, such as the thermodynamic uncertainty relation (TUR) \cite{Barato2015,Gingrich2016,MacIeszczak2018,Polettini2017,Pietzonka2017}, also generalized and obtained directly from the EFT \cite{Hasegawa2019,Timpanaro2019B}. In the tightest form, it reads $var(\mathcal{Q}_i)/\langle \mathcal{Q}_i \rangle^2 \geq f(\langle \Sigma \rangle)$, for some known function $f(x)$. From (\ref{DFT}), the underlying TUR is also valid for the entropy production itself, $var(\Sigma)/\langle \Sigma \rangle^2 \geq f(\langle \Sigma \rangle)$. Thus, TUR is seen as another bound concerning the statistics of $\Sigma$, such as the second law. For the TUR, the uncertainty of $\Sigma$ (and the currents $\mathcal{Q}_i$) is quantified in terms of the signal-to-noise ratio. 

In this context, it seems opportune to analyze the random variable $\Sigma$ with other tools that account for uncertainty, and a successful one comes from information theory \cite{Shannon1949,Cover}. After its debut, the theory  was readily recognized as of great importance to statistical mechanics but, in the words of Jaynes, ``the exact way in which it should be applied has remained obscure'' \cite{Jaynes1957}. Notable applications in physics were built in the works that followed \cite{Vedral2002,Peres2004,Adesso2019,Maruyama2009,Parrondo2015}. In particular, an important development was to recognize the Kullback-Leibler divergence \cite{KullbackS.andLeibler1951} (related to Shannon's entropy) as a Lyapunov function of Markov chains \cite{Schlogl1971,Germany1976}, a typical scenario found in the weak coupling approximation of thermodynamics.

In this paper, we use concepts of information theory to tackle the following problem: For a given mean $\langle \Sigma \rangle$, how much information, or surprise, should one expect in the distribution $P(\Sigma)$ that satisfies the DFT (\ref{DFT})? As it turns out, the information of $P(\Sigma)$ is upper bounded in terms of the mean, $\langle \Sigma \rangle$. More precisely, for a given discrete support $s=\{\Sigma_i\}$, we quantify the information of the entropy production in terms of its Shannon's entropy,
\begin{equation}
\label{info}
    H[\Sigma] := - \sum_{i} P(\Sigma_i) \ln P(\Sigma_i),
\end{equation}
here simply called information, where the sum is over $\Sigma_i \in s$. Then, we find a tight upper bound for (\ref{info}) from the DFT (\ref{DFT}), namely
\begin{equation}
\label{bounds}
H[\Sigma] \leq M(\langle \Sigma \rangle).
\end{equation}
The bound is given in terms of the information (\ref{info}) of the following maximal distribution:
\begin{equation}
\label{maximal}
    P_M(\Sigma)=\frac{1}{Z(\lambda)} \exp\Big({\frac{\Sigma}{2}-\lambda\frac{\Sigma}{2}\tanh{\frac{\Sigma}{2}}}\Big),
\end{equation}
defined over the discrete support $s$, which also can be written in terms of its mean, $M(\langle \Sigma \rangle)=\ln Z(\lambda) + (\lambda - 1)\langle \Sigma \rangle$/2, using the constraints $Z(\lambda)=\sum_i \exp(\Sigma_i/2 - \lambda \Sigma_i/2\tanh(\Sigma_i/2))$ and $-\partial_\lambda \ln Z(\lambda)=\langle \Sigma \rangle/2$. For continuous distributions $P(\Sigma)$, the upper bound also holds for differential entropy, $h[\Sigma]=-\int_{-\infty}^{\infty} P(\Sigma)\ln P(\Sigma)d\Sigma$, with full support in the real line. Note that the maximal distribution (\ref{maximal}) is a member of the exponential family \cite{RobertW.Keener2010}, but it has a nonlinear argument that seems unusual at first glance. As a matter of fact, we argue that this nonlinear structure is rather intuitive when combining the information maximization with the DFT (\ref{DFT}), as discussed below.

%\section{Formalism}
{\bf \emph{Formalism -}}
In this section, we find the upper bound for the information (\ref{info}) of the entropy production for a given mean. In this case, the DFT (\ref{DFT}) acts as a constraint. Alternatively, previous approaches \cite{Dewar2005} have found some form of FT from the MaxEnt procedure. In our paper, we start from a different point as we are interested in the impact of the DFT in the statistics of $\Sigma$ in the same sense of the derivation of the TUR.  We consider a general point mass function $P(\Sigma)$ over a discrete support, $s=\{\Sigma_i\} \subset \mathbb{R} $, i.e., $P(\Sigma)>0$ for all $\Sigma \in s$ and $P(\Sigma)=0$ otherwise. Without loss of generality, we consider $0\in s$. Additionally, $P(\Sigma)$ satisfies normalization, $\sum_{i} P(\Sigma_i) = 1$  and known mean $\sum_{i} \Sigma_i P(\Sigma_i) = \langle \Sigma \rangle$
where the summation is assumed over $s$. Finally, $P(\Sigma)$ also satisfies a detailed fluctuation theorem (\ref{DFT}) in $s$, which means the support is symmetric (for all $\Sigma_i\in s$, we have $-\Sigma_i \in s$). In the text, we use the terms distribution and point mass function (pmf) interchangeably.

First, define new variables $\sigma=sgn(\Sigma)$ for $\Sigma \neq 0$ and $\varepsilon=|\Sigma|$ with supports $\{-1,+1\}$ and $s\geq0$, respectively, and distributions $p(\sigma)$ and $q(\varepsilon)$. Using Bayes theorem, one has $ P(\Sigma)=p(\sigma|\varepsilon)q(\varepsilon)$,
where the fluctuation theorem (\ref{DFT}) defines $p(\sigma|\varepsilon)$ uniquely,
\begin{equation}
\label{psigma}
    p(\sigma|\varepsilon)=\frac{e^{\sigma \varepsilon/2}}{e^{\varepsilon/2}+e^{-\varepsilon/2}},
\end{equation}
for $\varepsilon > 0$ and $p(\sigma|0)=1$. Note that
$\langle \Sigma \rangle=\langle \sigma \varepsilon \rangle = \langle \varepsilon \tanh(\varepsilon/2)\rangle$, using (\ref{psigma}). Also note that $\Sigma \tanh (\Sigma/2)=\varepsilon\tanh(\varepsilon/2)$ by definition of $\varepsilon$, which leads to the following identity:
\begin{equation}
\label{identity}
    \langle \Sigma \rangle = \langle \Sigma \tanh(\Sigma/2)\rangle,
\end{equation}
that will be useful later, in analogy to similar treatments \cite{Hasegawa2019,Merhav2010}.  Now we find the upper bound of the information (\ref{info}) using calculus of variations. Usually, in the MaxEnt recipe, we have integral constraints (for instance, $\langle \Sigma \rangle$), but (\ref{DFT}) is not integral: It is a detailed relation that couples the negative and positive parts of the support. This symmetry allows the information (\ref{info}) to be written as a functional of $P(\Sigma)$ for $s>0:=\{\Sigma_i \in s |~\Sigma_i > 0\}$:
\begin{equation}
\label{variations1}
    H[\Sigma] = - \sum_{s>0} P(\Sigma)\big[\ln P(\Sigma)(1+e^{-\Sigma})-\Sigma e^{-\Sigma}\big]-P_0\ln P_0,
\end{equation}
for $P_0 = P(0)$. The same idea is applied to the constraints, resulting in the following functionals:
\begin{eqnarray}
\label{constraint1}
\sum_{s>0}\Sigma (1-e^{-\Sigma})P(\Sigma)=\langle \Sigma \rangle,
\\
\label{constraint2}
\sum_{s>0}(1+e^{-\Sigma})P(\Sigma)+P_0=1.
\end{eqnarray}
Finally, introducing Lagrange multipliers $\alpha$ and $\beta$ for both integral constraints (\ref{constraint1}) and (\ref{constraint2}), we obtain the maximization,
\begin{multline*}
\label{variations2}
\sum_{s>0} \delta P(\Sigma)\big[(1+e^{-\Sigma})\log P(\Sigma)-\Sigma e^{-\Sigma}+\\
  \alpha \Sigma (1-e^{-\Sigma})+(\beta+1)(1+e^{-\Sigma})\big]
  +\delta P_0 \{\log P_0 + \beta+1\}=0,
\end{multline*}
which is solved for 
$\ln P_0 = -\beta -1$ and
\begin{equation}
\label{maximal2}
    P_M(\Sigma)=P_0 \exp \big( \Sigma \frac{e^{-\Sigma}-\alpha(1-e^{-\Sigma})}{1+e^{-\Sigma}}\big).
\end{equation}
Redefining the parameters $\lambda = 2\alpha + 1$ and $P_0=1/Z(\lambda)$, we get the form (\ref{maximal}) valid for all $s$, as the negative part of the support is fixed by (\ref{DFT}), $P_M(-\Sigma)=e^{-\Sigma}P_M(\Sigma)$. The information (\ref{info}) for the maximal distribution (\ref{maximal}) is then the upper bound, which is our main result,
\begin{equation}
\label{upperbound}
    H[\Sigma]\leq M(\langle \Sigma \rangle) := \ln Z(\lambda) +\frac{\langle \Sigma \rangle }{2}(\lambda-1),
\end{equation}
where (\ref{identity}) was used explicitly to write the upper bound $M(\langle \Sigma \rangle)$ in terms of the mean $\langle \Sigma \rangle$, for $Z(\lambda)=\sum_i \exp(\Sigma_i/2 - \lambda \Sigma_i/2\tanh(\Sigma_i/2))$ and $-\partial_\lambda \ln Z(\lambda)=\langle \Sigma/2 \tanh (\Sigma/2)\rangle=\langle \Sigma \rangle/2$, also from (\ref{identity}), proving the upper bound in (\ref{bounds}). For the case where $0 \notin s$, then $P_0=0$ and $P(\Sigma)$ is still given by (\ref{maximal}) for $\Sigma \neq 0$ with $Z(\lambda)$ defined accordingly.

The derivation of the upper bound for the continuous case ($s=\mathbb{R}$) is straightforward if one uses the differential entropy
\begin{equation}
\label{differential}
    h[\Sigma]=-\int_{-\infty}^\infty P(\Sigma)\ln P(\Sigma)d\Sigma
\end{equation}
with suitable constraints $\int P(\Sigma)d\Sigma=\int_{\Sigma>0} P(\Sigma)(1+\exp(-\Sigma))d\Sigma=1$, $\int \Sigma P(\Sigma)d\Sigma=\int_{\Sigma>0} \Sigma(1-\exp(-\Sigma))P(\Sigma)d\Sigma=\langle \Sigma \rangle$ and repeating steps (\ref{variations1})-(\ref{upperbound}), we get
\begin{equation}
\label{differentialbound}
h[\Sigma] \leq m(\langle \Sigma\rangle):=\ln Z(\lambda) +\frac{\langle \Sigma \rangle }{2}(\lambda-1).
\end{equation}
In this case, the maximal distribution (\ref{maximal}) is defined for the real line with $Z(\lambda)=\int \exp(\Sigma/2-\lambda \Sigma/2 \tanh(\Sigma/2))d\Sigma$ and $-\partial_ \lambda \ln Z(\lambda)=\langle \Sigma \rangle/2$ defining $\lambda$ and $Z(\lambda)$ implicitly. A way to check inequality (\ref{differentialbound}) directly is through Gibbs' inequality. Define the Kullback-Leibler divergence $D(P||P_M)=\int P(\Sigma)\ln(P(\Sigma)/P_M(\Sigma))d\Sigma \geq 0$, which implies, for this case, $D(P|P_M)=m(\langle \Sigma \rangle)-h[\Sigma] \geq 0$.

%In the choice of the measure in (\ref{differential}), notice that we have a particular case of entropy production from the EFT, where $\Sigma = A Q = A(E_2 - E_1)$ (in one dimension, for simplicity). In this particular case, the definition of $\Sigma$ works as a transformation of random variables from ($E_1,E_2$) to ($E_1,\Sigma$), where $d\Sigma = A dE_2$. Intuitively, in the EFT, it seems natural that the entropy production naturally uses the measure from the energy density, although the result could be generalized with respect to other reference measures.

We argue that the nonlinear term, $\Sigma \tanh \Sigma/2$, of the maximal distribution (\ref{maximal}) becomes intuitive after learning identity (\ref{identity}). In the generalized Gibbsian ensemble (GGE), the general solution of MaxEnt problems, we get the exponents from the constraints. As an example, the famous derivation of the Boltzmann weights, $P(E_i) \propto e^{-\beta E_i}$, from constraint $\langle E_i \rangle$. In our case, as the constraint $\langle \Sigma \rangle$ is augmented to an extra constraint in $\langle \Sigma \tanh(\Sigma/2) \rangle$, due to identity (\ref{identity}), it is intuitive that both forms appear in the exponent of the maximal distribution (\ref{maximal}). Actually, normalization is also augmented to $\langle 1 \rangle = \langle \coth(\Sigma/2)\rangle=1$, however, as the DFT (\ref{DFT}) is stronger than (\ref{identity}), it fixes the odd term in the exponent of $P(\Sigma)$ to exactly $\Sigma/2$. In the following sections, we check the upper bounds in the discrete (\ref{upperbound}) and continuous (\ref{differentialbound}) cases for different relevant physical systems.

\begin{figure}[htp]
\includegraphics[width=3.3 in]{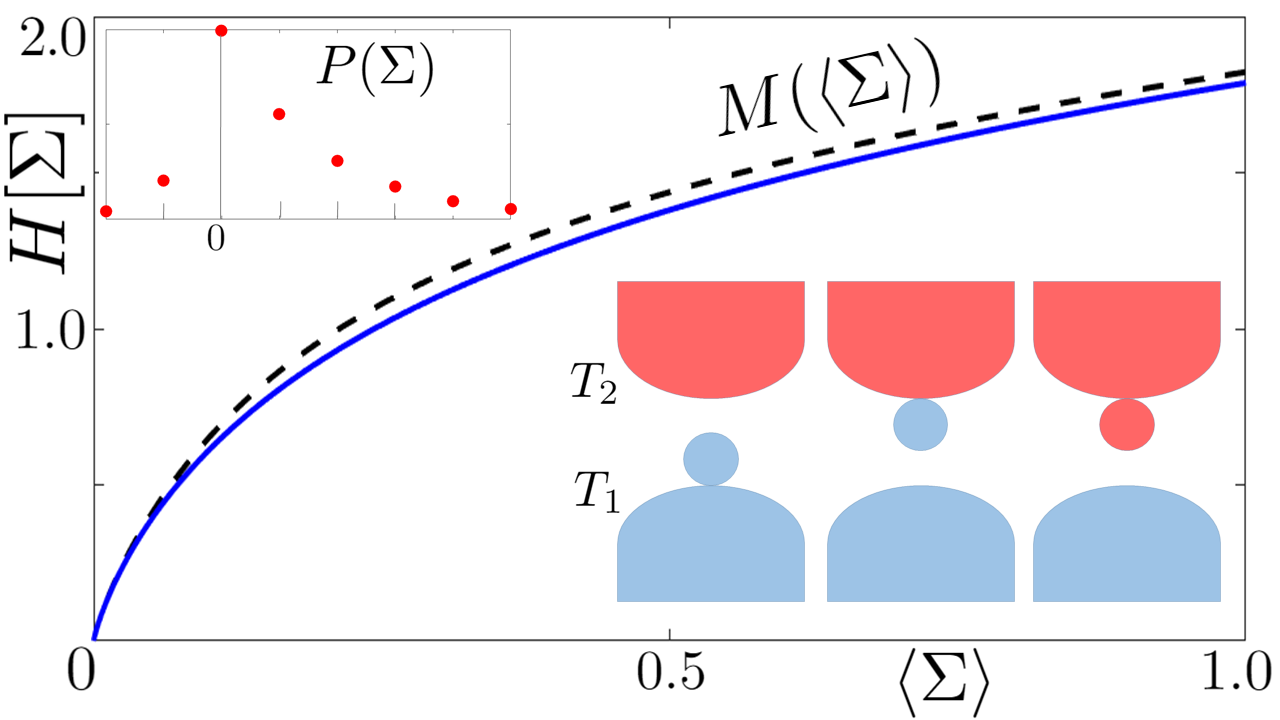}
\caption{(Color online) The information of the entropy production $H[\Sigma]$ as a function of the mean, $\langle \Sigma \rangle$, for the bosonic mode at weak coupling (blue line) and the theoretical upper bound (dashed) given by $M(\langle \Sigma \rangle)$ slightly above it. The  system is prepared in thermal equilibrium with the first reservoir at temperature $T_1$ and, at $t=0$, the system is coupled to the second reservoir at temperature $T_2$ (lower inset). A two-point measurement scheme takes places for $t=0$ and $t>0$, and the entropy production is given by $\Sigma=\Delta \beta \Delta E$. The pmf for $\langle \Sigma \rangle=1$ is depicted (upper inset).}
\label{fig1}
\end{figure}
%\section{Applications}
{\bf \emph{Application to a bosonic mode -}} Consider a bosonic mode with Hamiltonian $H=\hbar\omega (a^\dagger a+1/2)$ weakly coupled to a thermal bath such that its dynamics satisfies a Lindblad's equation \cite{Santos2017a,Salazar2019,Denzler2019},
\begin{equation}
\label{Lind}
\partial_t \rho = \frac{-i}{\hbar}[H,\rho] + D_i(\rho)
\end{equation}
for the dissipator given by
\begin{equation}
D_i(\rho)=\gamma(\overline{n}_i+1)[a\rho a^\dagger - \frac{1}{2}\{a^\dagger a, \rho\}]+\gamma \overline{n}_i[a^\dagger \rho a -\frac{1}{2}\{a a^\dagger , \rho\}],
\end{equation}
where $\gamma$ is the dissipation constant and $\overline{n}_i=[\exp(\hbar \omega/k_BT_i)-1]^{-1}$ is the bosonic thermal occupation number and $\beta_i=1/(k_b T_i)$. The system is prepared in equilibrium with temperature $T_1$ and at $t=0$ it is placed in thermal contact with the second reservoir (at temperature $T_2$). Using a two point measurement scheme (at $t=0$ and $t>0$) in the absence of any external protocol, the entropy production is given in terms of the energy variation \cite{Campisi2015,Timpanaro2019B,Sinitsyn2011} as $\Sigma = -(\beta_2-\beta_1)\Delta E$. This setup maps thermal initial states in time dependent thermal states, for some time dependent temperature satisfying a ``law of cooling'' (from $T_1$ to $T_2$) \cite{Salazar2019}. Moreover, this property allows the nonequilibrium heat distribution (and the entropy production) to be written solely in terms of the equilibrium partition function and the ``law of cooling''. Using this idea, the distribution $P(\Sigma)$ follows \cite{Salazar2019,Denzler2019}  directly:
\begin{equation}
\label{PsigmaHO}
    P(\Sigma)=\frac{1}{A(\alpha)} \exp(\frac{\Sigma}{2}-\alpha \frac{|\Sigma|}{2}),
\end{equation}
with support $s=\{\pm \Delta \beta \hbar \omega m\}$, $m=0,1,2,..$, and constants $A=A(\alpha)$ and $\alpha$ uniquely defined from the normalization and mean constraints. Note that (\ref{PsigmaHO}) satisfies (\ref{DFT}). The information (\ref{info}) of the distribution (\ref{PsigmaHO}) is given by
\begin{equation}
\label{QHOentropy}
H[\Sigma]= \ln A(\alpha) -\frac{\langle \Sigma \rangle}{2} + \alpha \frac{\langle |\Sigma| \rangle}{2},
\end{equation}

\begin{figure}[htp]
\includegraphics[width=3.3 in]{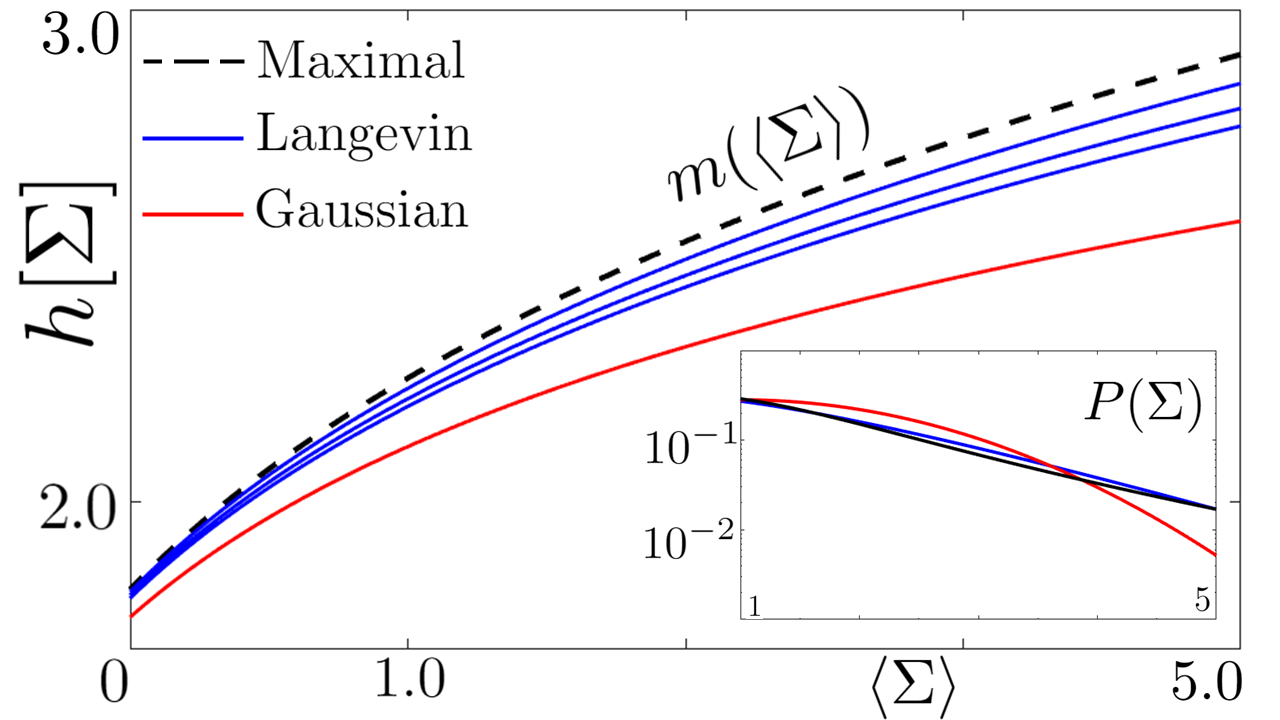}
\caption{(Color online) The information of the entropy production, now measured by the continuous $h[\Sigma]$, as a function of the mean, $\langle \Sigma \rangle$, for the Langevin model for a levitated nanoparticle (blue curves, $d=1,2,3$, respectively from top to bottom); theoretical upper bound (dashed) given by $m(\langle \Sigma \rangle)$ and Gaussian distribution information (red). The tails of the distributions are depicted in the inset (for $\langle \Sigma \rangle=1$) and the approximately exponential decay of the maximal and Langevin cases ($d=1$) is explicit.
}
\label{fig2}
\end{figure}
where $\langle | \Sigma | \rangle$ may be written in closed form in terms of $\alpha$ using the geometric series. In Fig.~1, we compute the information (\ref{QHOentropy}) and the mean $\langle \Sigma \rangle$ for several values of $\alpha$ for the distribution (\ref{PsigmaHO}). Instead of plotting $H[\Sigma]$ as a function of $\alpha$, we plot $H[\Sigma]$ vs. $\langle \Sigma \rangle$, yielding a single blue curve. We repeat the same process with the maximal distribution (\ref{maximal}), computing its information and mean $\langle \Sigma \rangle$ for several values of $\lambda$, then we also plot $M(\langle\Sigma\rangle)$ vs. $\langle \Sigma \rangle$ (single dashed curve). In Fig.1, notice that the upper bound is always above the information of the system's entropy production by a small amount. 

Actually, the entropy production in this case has the general form $\Sigma = -\Delta \beta \Delta E = AQ$, where $A=\Delta \beta$ and $Q=\pm \hbar \omega m$. For the limiting case, $\hbar \omega \Delta \beta \gg 1$, one has $|\pm \Sigma_m|=m (\hbar \omega \Delta \beta)\gg1 $ 
and the following approximation holds
\begin{equation}
    \frac{ \Sigma_m}{2}\tanh{\frac{\Sigma_m}{2}}
    \approx \frac{|\Sigma_m|}{2},
\end{equation}
valid for $\pm \Sigma_m$, which makes the exponent of the maximal distribution (\ref{maximal}) similar to the observed (\ref{PsigmaHO}), $ P_M(\Sigma)\approx 1/Z(\lambda)\exp(\Sigma/2-\lambda|\Sigma|/2)$. We conclude that, depending on the support, i.e., the interplay between quantum energy levels and affinities, the maximal distribution is approximately attained for entropy production in the heat transfer using a bosonic mode at weak coupling. 

{\bf \emph{Application to a Gaussian distribution -}} The Gaussian distribution has a broad range of applications also in the context of entropy production \cite{Pigolotti2017,Chun2019}. The DFT (\ref{DFT}) allows one to write its standard deviation as a function of the mean, and the resulting pdf is
\begin{equation}
\label{gaussian}
    P(\Sigma)=\frac{1}{2\sqrt{\pi \langle \Sigma \rangle}}\exp\Big(\frac{-(\Sigma-\langle \Sigma\rangle)^2}{4\langle \Sigma \rangle}\Big),
\end{equation}
where it clearly satisfies (\ref{DFT}), $\int P(\Sigma)d\Sigma=1$ and $\int \Sigma P(\Sigma)d\Sigma=\langle \Sigma \rangle$. Therefore, the differential entropy (\ref{differential}) for the Gaussian case (\ref{gaussian}) must satisfy the upper bound, 
\begin{equation}
\label{gaussianentropy}
   h[\Sigma]=\frac{1}{2}\ln(4\pi e \langle \Sigma \rangle)\leq m(\langle \Sigma \rangle),
\end{equation}
which is a general inequality for the function $m(\langle \Sigma \rangle)$ defined in (\ref{differentialbound}). This inequality is depicted in Fig.~2.

{\bf \emph{Application to a levitated nanoparticle -}} The highly underdamped limit of the Langevin equation represents the dynamics of a levitated nanoparticle \cite{Gieseler2012,Gieseler2018,Aspelmeyer2014}. Consider the Langevin dynamics with potential $\mathcal{U}(x)=m k x^{2}/2$ in one dimension for simplicity. The particle's dynamics is given by
\begin{equation}
    \label{Langevin}
    \ddot{x} + \Gamma \dot{x} +\Omega_0^2 x = \frac{1}{m}F_{fluc}(t),
\end{equation}
for position $x(t)$, with Gaussian noise $\langle F_{fluc}(t)F_{fluc}(t')\rangle = 2m\Gamma T \delta(t-t')$, where $\Gamma$ is a friction coefficient, $m=1$ is the particle mass, $T$ is the reservoir temperature and $k=\Omega_0^2$ is a constant (not driven by a protocol). Define a $d-$dimensional system energy $E=\sum_i^d [p_i^2/2+\mathcal{U}(x_i)]$, with momentum $p_i=\dot{x_i}$; the following stochastic differential equation (SDE) was obtained for the total energy in the highly underdamped limit \cite{Gieseler2012}, $\Omega_0\gg\Gamma$:
\begin{equation}
\label{LangevinforE2}
dE=-\Gamma (E- \frac{f}{2}T)dt+\sqrt{2\Gamma TE}d\textrm{W}_t,
\end{equation}
with degrees of freedom $f=2d$ and  $d\textrm{W}_t$ is a Wiener increment. Using the same setup of Fig.~1, but now with the levitated nanoparticle as working medium for the heat transfer, one defines the entropy production as $\Sigma = -\Delta \beta \Delta E$, where $P_t(\Delta E)=\int P(E_1)\Pi_t(E_1\rightarrow E_2)\delta (\Delta E - (E_2-E_1))dE_1 dE_2$. The propagator $\Pi_t$ is known \cite{Gieseler2018,Salazar2016} for the SDE (\ref{LangevinforE2}), from the solution of its Fokker-Planck equation, which yields the following distribution:
\begin{equation}
\label{nanopdf}
    P(\Sigma)=\frac{1}{B(\alpha)}
    \exp(\frac{\Sigma}{2})|\Sigma|^{d+1/2}K_{d+1/2}(\alpha |\Sigma|),
\end{equation}
defined for the real line for constant $\alpha$ defined in terms of parameters ($T_1,T_2,\Gamma t$), $B(\alpha)$ is a normalization constant, and $K$ is the modified Bessel function of the second kind. In Fig.~2, the information $H[\Sigma]$ and the mean $\langle \Sigma \rangle$ of the distribution (\ref{nanopdf}) are numerically computed for several values of $\alpha$ and different sizes ($d=1-3$). For each value of $\alpha$, we plot the blue curves $H[\Sigma]$ vs $\langle \Sigma \rangle$, one for each $d=1-3$. The same process is repeated for the upper bound (\ref{differentialbound}), resulting in the dashed curve. For comparison the Gaussian distribution was included (red curve), using (\ref{gaussianentropy}). Inspecting Fig.2, one sees that the observed entropy production is close to the upper bound, especially for the case of $d=1$, also with good agreement in the tails (inset). Larger systems ($d=2,3$) and the Gaussian case have lower information and misses the bound by a larger amount for $\langle \Sigma \rangle \gg 1$.

{\bf \emph{Application to swap engines -}} We consider a pair of qubits with energy gaps $\epsilon_A$ and $\epsilon_B$ initially prepared in thermal equilibrium, $p(\sigma)=\exp(\sigma\beta \epsilon)/(\exp(-\beta\epsilon)+\exp(+\beta\epsilon))$, for $\sigma=\pm1$, $\beta\in\{\beta_1,\beta_2\}$ and $\epsilon\in\{\epsilon_A,\epsilon_B\}$, with reservoirs at temperature $T_1$ and $T_2$, respectively. A two point energy measurement is performed before and after a swap operation \cite{Campisi2015} takes place, 
\begin{figure}[htp]
\includegraphics[width=3.3 in]{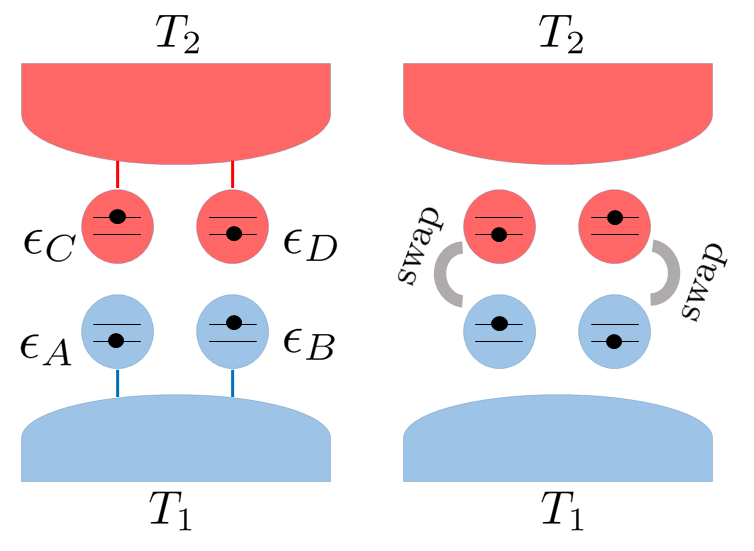}
\caption{(Color online) Four qubits are placed in thermal equilibrium with two reservoirs. The pairs $(A,C)$ and $(B,D)$ undergo a swap operation as depicted, which produces entropy obtained in terms of energies of a two point measurement scheme (before and after the swap). The system is equivalent to a composition of two independent swap engines and, with suitable choices of energy gaps (for instance, $\epsilon_C/\epsilon_A=\epsilon_D/\epsilon_B$, $\beta_1/\beta_2=1/2$, $\epsilon_A/\epsilon_B=\epsilon_C/\epsilon_D=2/3$), the entropy production after the swap assumes nine possible outcomes $\{0,\pm b, \pm 2b, \pm 3b, \pm 5b\}$ for a constant $b$ instead of three outcomes observed in the single pair swap engine. We show that, even for this nontrivial support, the entropy production distribution is a particular case of the maximal distribution.
}
\label{fig3}
\end{figure}
defined as $|xy\rangle \rightarrow |yx\rangle$, for $x,y \in \{-,+\}$ and the entropy production in the process is given \cite{Campisi2015,Timpanaro2019B} by
\begin{equation}
\label{swapprod}
    \Sigma = \beta_1 \Delta E_A + \beta_2 \Delta E_B,
\end{equation}
where $\Delta E_A = E_A^f-E_A^i$, $\Delta E_B=E_B^f-E_B^i$ are the variations of energy measurements before and after the swap. Therefore, in this measurement scheme, the three possible outcomes are $\Sigma \in s=\{0,\pm 2a\}$ for $2a=2(\beta_2\epsilon_B-\beta_1\epsilon_A)$. The distribution $P(\Sigma)$ for the swap operation follows from initial state distributions directly:
\begin{equation}
    \label{swapdist}
    P(\Sigma)=\frac{1}{Z_0}\exp\Big(\frac{\Sigma}{2}\Big),
\end{equation}
for $\Sigma \in s$, which satisfies the DFT (\ref{DFT}) and it is a particular case of the maximal distribution (\ref{maximal}) for $\lambda=0$. In this case, the information reads $H[\Sigma]=M(\langle \Sigma \rangle)=\ln Z_0 - \langle \Sigma \rangle/2,$ for $Z_0=\sum_i \exp(\Sigma_i/2)$. This is a trivial example of the maximal distribution because any distribution satisfying the DFT in a support $s$ with only three values always has the form (\ref{swapdist}).

However, we show that larger swap engines preserve the form (\ref{swapdist}) with nontrivial supports. For instance, consider the double swap engine formed by four qubits with energy gaps $(\epsilon_A,\epsilon_B,\epsilon_C,\epsilon_D)$ arranged as depicted in Fig.~3. Qubits $A$ and $B$ ($C$ and $D$) are in thermal equilibrium with a reservoir at temperature $T_1$ ($T_2$). A swap operation takes place between qubits $A$ and $C$ (pair 1). Simultaneously, another swap is performed with qubits $B$ and $D$ (pair 2). We choose the energy gaps such that $r:=\epsilon_A/\epsilon_C=\epsilon_B/\epsilon_D$, i.e., the independent engines ($A,C$) and ($B,D$) operate similarly. Additionally, the independent engines are related by $\epsilon_A/\epsilon_B=2/3$. For simplicity, let $\beta_1/\beta_2=1/2$. The entropy production now is given by
\begin{equation}
\label{swapprod2}
    \Sigma = \beta_1 (\Delta E_A+\Delta E_B) + \beta_2 (\Delta E_C+\Delta E_D)=\Sigma_1+\Sigma_2,
\end{equation}
where $\Sigma_i$ is the entropy production of the independent pair $i=1,2$ (\ref{swapprod}). One can easily check that the supports of $\Sigma_1$ and $\Sigma_2$ are $s_1=\{0,\pm 2b\}$ and $s_2=\{0,\pm 3b\}$, and their composition results in nine different outcomes for (\ref{swapprod2}) $s=\{0,\pm b, \pm 2b, \pm 3b, \pm 5b\}$, all multiples of $b=\beta_2 \epsilon_C (1-r/2)$. In this specific case, the distribution is also given by (\ref{swapdist}), which follows from $P(\Sigma=\Sigma_1+\Sigma_2)=P_1(\Sigma_1)P_2(\Sigma_2)=\exp(\Sigma_1/2)C_1  \exp(\Sigma_2/2)C_2=C \exp(\Sigma/2)$, which is a maximal distribution. To summarize, a composite microscopic swap engine, now with nine possible outcomes in the support of the entropy production, still behaves as a particular case of the maximal distribution. This is particularly interesting since the swap operation is the optimal unitary operation that  outputs the most work per cycle \cite{Campisi2015}. The argument is easily generalized for larger compositions of swap engines, for suitable choices of energy gaps.

{\bf \emph{Other applications -}} It is worth noting that the strong DFT (\ref{DFT}) also holds for deterministic dynamical ensembles \cite{Hasegawa2019,Evans1993}. In this case, one has $N$ particles described by a deterministic trajectory in the phase space, where randomness is encoded in the initial distribution. It was proved that, for some assumptions in the distribution and dynamics, the system satisfies (\ref{DFT}). Therefore, the upper bound (\ref{differentialbound}) is expected to hold for such systems.

%\section{Discussion and Conclusions}
{\bf \emph{Conclusions -}} In this paper, we used information to quantify the uncertainty in the entropy production. We obtained an upper tight bound for a given mean in terms of a proposed maximal distribution, $P_M(\Sigma)$. We argued that the non-linearity observed in $P_M(\Sigma)$ is a result of a symmetry derived from the DFT. Then, we verified the behavior of some relevant distributions in comparison to the maximal. Namely, transferring heat between two reservoirs using a bosonic mode results in a distribution close to $P_M(\Sigma)$, specially in a limiting case. In the same setup, a levitated nanoparticle yielded a distribution close to the maximal, but now in the continuous domain. For the composite swap qubit engine, we found that a case of the maximal distribution is always observed. In this context, analyzing the role of mutual information to quantify dependencies between thermodynamic variables is left for future research.

We remark that our main result falls in the same category of the TUR \cite{Timpanaro2019B}: A bound for the statistics of $\Sigma$ derived from the fluctuation theorem. Both bounds are only saturated for very specific systems. In general, the underlying mechanisms of the nonequilibrium dynamics will likely introduce additional constraints in the entropy production and the maximal distribution will not be observed. This is fundamentally different from the MaxEnt derivation in equilibrium thermodynamics, where the Maxwell-Boltzmann distribution not only bounds the thermodynamic entropy but saturation of the bound is also expected for systems in equilibrium. 

\bibliography{library}
\end{document}